\documentclass[journal=jacsat,manuscript=article,layout=twocolumn]{achemso}
\usepackage[version=3]{mhchem} %
\usepackage[T1]{fontenc}       %
\usepackage{array,multirow,graphicx}
\usepackage{xspace}
\usepackage{xr,textcomp}

\newcommand{\mg}{Mg$^{2+}$\xspace}
\newcommand{\kp}{K$^{+}$\xspace}
\newcommand{\cl}{Cl$^{-}$\xspace}
\newcommand*\rot{\rotatebox{90}}
\newcolumntype{L}[1]{>{\raggedright\let\newline\\\arraybackslash\hspace{0pt}}m{#1}}
\newcolumntype{C}[1]{>{\centering\let\newline\\\arraybackslash\hspace{0pt}}m{#1}}
\newcolumntype{R}[1]{>{\raggedleft\let\newline\\\arraybackslash\hspace{0pt}}m{#1}}

\author{Richard A. Cunha}
\affiliation[SISSA]
{International School for Advanced Studies, Trieste, Italy}
\author{Giovanni Bussi}
\affiliation[SISSA]{International School for Advanced Studies, Trieste, Italy}
\email{bussi@sissa.it}
\phone{+39 040 3787 407 }

\title[]
  {
    Unravelling Mg$^{2+}$-RNA binding with atomistic molecular dynamics
  }

\abbreviations{MD,MetaD,RNA}
\keywords{Flexible RNA, Magnesium Ions,Hybridization}

\begin{document}

\newcommand*{\sometext}{
Interaction with divalent cations is of paramount importance for RNA structural stability and function. We here report a detailed molecular dynamics study
of all the possible binding sites for \mg on a RNA duplex, including both direct (inner sphere) and indirect (outer sphere)  binding.
In order to tackle sampling issues, we develop a modified version of bias-exchange metadynamics
which allows us to simultaneously compute affinities with previously unreported statistical accuracy.
Results correctly reproduce trends observed in crystallographic databases.
Based on this, we simulate a carefully chosen set of models that allows us to quantify the effects of competition with monovalent cations,
RNA flexibility, and RNA hybridization.
Our simulations reproduce the decrease and increase of \mg affinity  due to ion competition and hybridization respectively,
and predict that RNA flexibility has a site dependent effect.
This suggests a non trivial interplay between RNA conformational entropy and divalent cation binding.

 }
\let\oldmaketitle\maketitle
\let\maketitle\relax

\makeatother

\noindent \twocolumn[
 \begin{@twocolumnfalse}
   \oldmaketitle
   \begin{abstract}
     \sometext
     \leavevmode \\
   \end{abstract}
 \end{@twocolumnfalse}
 ]

\section{Introduction}

The relevance of ribonucleic acids (RNA) in molecular biology has systematically grown since the
discovery that it can catalyze chemical reactions \cite{doudna2002repertoireribozymes}
and RNA is now considered a key player in many of the regulatory networks of the cell.\cite{morris2014riseregRNA}
Functions such as catalysis and regulation of gene expression rely on the peculiar structure and dynamics of RNA molecules.
The folding of RNA three-dimensional structure stands in a delicate balance between canonical interactions and strong electrostatics mediated by the presence of cations.
Cations, together with water, are indeed crucial to compensate for the large repulsion between the charged phosphate groups present in the RNA backbone.
They allow for the formation of tertiary contacts, \cite{tinoco1999howrnafolds,lipfert2014understanding}
and can also provide entropic stabilization to RNA motifs.\cite{fiore2012entropic}
Among  cations, Mg$^{2+}$ is particularly relevant because of its double charge and small radius. \cite{oliva2009frequency}
Mg$^{2+}$ can be both directly and indirectly bound to RNA, 
that is RNA atoms can be part of \mg inner coordination sphere or interact through hydrogen bonds with its hydration sphere. \cite{williams2012cations}
The inner sphere ions mainly contribute to the formation of specific structural motifs. \cite{Petrov2011clamps}.
The outer sphere ones might also bind to specific motifs and additionally take part to the ion atmosphere and stabilize RNA structures by screening electrostatic repulsion \cite{draper2004guidetoions}.
Several experimental works have provided valuable insights on the thermodynamics of RNA-Mg$^{2+}$ interactions in solution. \cite{Kirmizialtin2012scatteringexp,Draper2008rnathermodynamics,Bizarro2012forcespec,Erat2011methods}
Titration experiments have been used to characterize the overall affinity of RNA for Mg$^{2+}$. \cite{sigel2010titrations} 
Affinities for individual Mg$^{2+}$ binding sites on
RNA nucleosides and small RNA motifs have also been reported. \cite{Freisinger2007metalbinding} 
The role of functional metal ions on small RNAs have also been investigated through nuclear magnetic resonance spectroscopy.\cite{Hoffmann2003nmrmg,campbell2006nmrmgloop,Bartova2016nmr}
However, the precise characterization of typical Mg$^{2+}$ binding sites in large RNA molecules
has largely been obtained by analyzing crystal structures.\cite{banatao2003microenvironment,Auffinger2011metal} A recent database survey
allowed for a classification of all the binding modes observed in crystallographic structures. \cite{bujnicki2015pdbfreq}
Molecular modeling could in principle provide a powerful tool to bridge the gap between detailed crystallographic
structures and solution experiments. \cite{schlick2010molecular}
In this respect, several works at different level of resolution have been reported, ranging from quantum-chemistry calculations, \cite{Kolev2013abinitio,Tongraar2001mgqmcalc,mlynsky2015roleofmg,petrov2011bidentate,Casalino2016}
to explicit solvent molecular dynamics (MD), \cite{allner2012magnesium,Kirmizialtin2012mdarnamg,Pan2014iondistributions}
implicit solvent methods, \cite{hayes2015magnesium,mak2016implicitmg} and coarse-grained models.\cite{thirumalai2011coarsegrained}
Among the computational methods, MD presents an excellent compromise between accuracy and computational cost,
though the development of an appropriate parametrization for Mg$^{2+}$-RNA interactions is still a debated topic \cite{panteva2015comparison,bergonzo2016divalent}. 
Moreover, due to the high energetic barriers involved in Mg$^{2+}$-RNA and Mg$^{2+}$-water interaction \cite{bleuzen1997waterexch},
which brings the lifetime of these complexes to the ms timescale, coupling of MD with enhanced sampling methods is required.

In this Paper, we use a unique combination of enhanced sampling techniques together with a recently published parametrization~\cite{allner2012magnesium}
for Mg$^{2+}$ to compute its affinity on a flexible RNA duplex. RNA duplexes are the most recurring motifs observed in ribosomal RNA. \cite{Gutell1994ribosome}
The computed affinities for all the relevant binding sites are compared with previously reported thermodynamic data and with an analysis of the protein databank (PDB).
Furthermore, by performing simulations on an appropriately chosen set of model systems we are able to investigate the interplay between Mg$^{2+}$-RNA binding affinity and competition with monovalent ions, RNA flexibility, and RNA hybridization.

\section{Methods}

\begin{figure}[h]
  \includegraphics{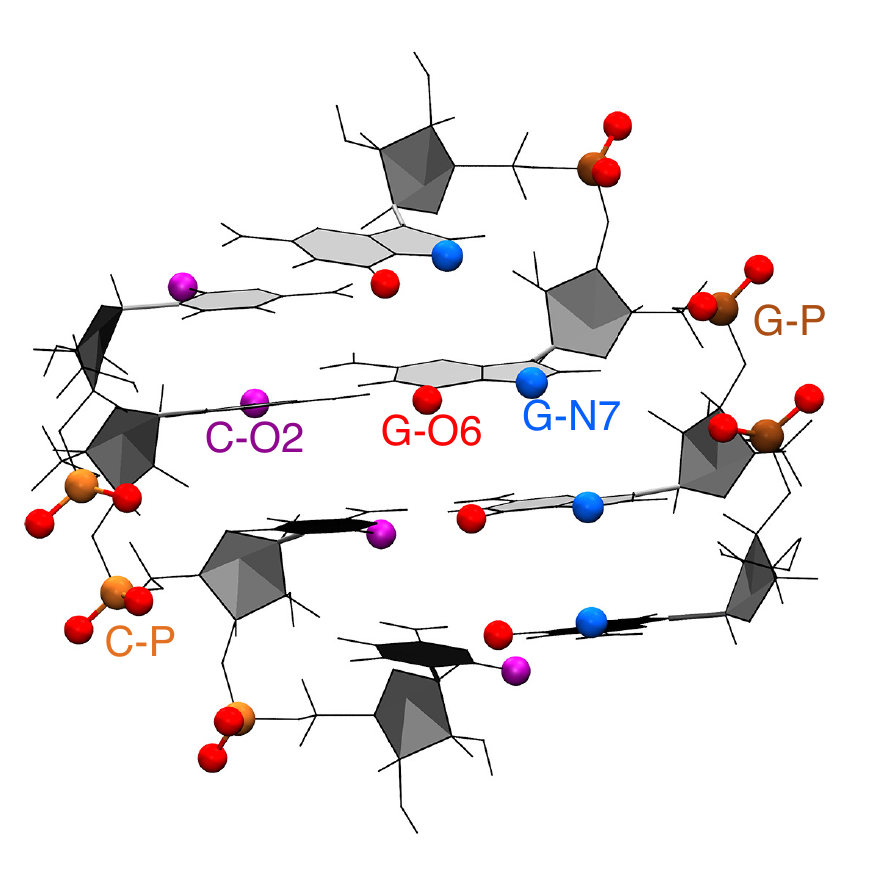}
  \caption{A-form RNA duplex with sequence $^\text{GGGG}_\text{CCCC}$. Target \mg binding sites are highlighted.
   }
  \label{fgr:duplex}
\end{figure}

The initial RNA structure was generated using the {\it make-na} webserver \cite{macke1998modeling} as an ideal A-form helix 
with sequence $^\text{GGGG}_\text{CCCC}$ (see Figure \ref{fgr:duplex}).
Molecular dynamics (MD) simulations of the duplex were performed using GROMACS 4.6.7. \cite{pronk2013gromacs}
RNA was described using the AMBER-ff99 force field with parmbsc0 and $\chi$OL corrections.\cite{cornell1995amber,perez2007parmbsc0,zgarbova2011refinement}
These parameters are available at http://github.com/srnas/ff.
The modeling of the monovalent ions (\kp and \cl), was done using the parameters proposed by Cheatham and cooworkers.\cite{cheatham2008alkali}
For Mg$^{2+}$ we used a parametrization developed in ref. \cite{allner2012magnesium}, which is also discussed further below. 
The duplex was solvated in a truncated dodecahedral box filled with explicit TIP3P water molecules. \cite{jorgensen1983tip3p}
Ions were added by substituting randomly selected water molecules.
Bonds were constrained using the LINCS algorithm, and the integration of the equations of motion were performed with a 2 fs timestep.
The temperature was set to 300K, and it was kept constant by a stochastic velocity rescale thermostat\cite{bussi2007canonical}.
Non-bonded interactions were calculated using the Verlet cutoff scheme, and electrostatics using particle-mesh Ewald. \cite{darden1993pme}
The cutoff was initially set to 1 nm and is adjusted adaptively so as to balance the load of real-space and reciprocal-space contributions.
Pressure was kept constant at 1bar using the Berendsen barostat for the equilibration phase\cite{berendsen1984molecular} and
the Parrinello-Rahman scheme during the production runs.\cite{parrinello1981barostat}
Simulations with a rigid duplex had the RNA atomic positions kept fixed, and were performed at constant volume.

Enhanced sampling simulations were then performed combining GROMACS with PLUMED 2.2.\cite{tribello2014plumed} 
A modified version of bias-exchange metadynamics (BE-MetaD)\cite{piana2007bemetad} was used to sample all the possible sites for inner sphere Mg$^{2+}$ binding.
In each replica we applied a bias potential according to MetaD in its well-tempered formulation,\cite{laio2002metad,barducci2008wtmetad} acting simultaneously on two collective variables, namely coordination number with water  ($CN_W$) and distance $d_i$ between Mg$^{2+}$ and $i$-th target binding sites, resulting in as many replicas as potential binding sites. 
A sample bidimensional free-energy surface profile is shown on SI 1.
In this work we identify the possible binding sites for both inner and outer sphere binding using the name of the
corresponding ligand (as highlighted in Figure \ref{fgr:duplex}).
To this aim, we only considered binding sites with an expected affinity large enough to require enhanced sampling (phosphates, O6 and N7 in guanine, O2 in cytosine, see Figure \ref{fgr:duplex}).  
We chose these sites based on preliminary simulations that we performed for all 4 RNA nucleosides and a guanine dinucleoside monophosphate
and also on available experimental information. \cite{bujnicki2015pdbfreq,sigel2010titrations}
This procedure resulted in 18 replicas and a total simulation time of 9 $\mu$s (18 replicas $\times$ $0.5\mu$s).
With the exception of a control simulation all the MD runs contained only one \mg ion, which is the one being biased by the MetaD.
In the control simulation including multiple \mg ions only one of them was biased. The coordination number of the remaining \mg was
restrained so as to avoid them to directly bind RNA.
In simulations performed with a larger box to study RNA hybridization the total time was 3.6 $\mu$s (18 replicas $\times$ $0.2\mu$s).
On the table \ref{simsystems} it is summarized the number of atoms, number of water molecules, number of ions and lenght for each simulation.
Additionally, penalty bias potentials were added to avoid the competition of different binding sites in the same replica.
This is not usually done in BE-MetaD but was required here to avoid Mg$^{2+}$ to be trapped in unbiased binding sites.
We also tried using the conventional BE-MetaD approach, but the sampling was undermined due to the Mg$^{2+}$ being stuck in other positions.
To ensure that the RNA helical structure was maintained through the enhanced sampling simulations, restraints were added to the distance and angles of all the hydrogen bonds corresponding to the four Watson-Crick base pairs.
All the replicas were run simultaneously, and the acceptance rate was calculated taking into account the bias potentials
introduced by the MetaD and penalty potentials on the unbiased binding sites.
All the replicas have the same temperature, and the difference between the ensembles comes from the bias introduced by the MetaD and from the penalty potentials,
which are unique for each replica. The acceptance for an exchange between replica $i$ and replica $j$ is thus evaluated as
$$
\alpha=\min\left(1,e^{-\frac{V_i(q_j)+V_j(q_i)-V_i(q_i)-V_j(q_j)}{k_B T}  }\right)
$$
Here $V_i$ is the bias potential acting on replica $i$, including both the MetaD potential and the penalty bias potential, and $q_i$ are the coordinates
for the replica $i$.
Ergodicity was thus ensured by accelerating the binding and unbinding events on all the possible binding sites with significant free-energy barriers.
A sample input file for one of the model systems is provided in SI 2.

\begin{table*}[t]\footnotesize
  \caption{Table containing the list of the studied systems, its components and the total simulation time. The simulations vary in the following parameters: small box ($sb$) vs large box ($lb$), rigid (fixed RNA atomic positions) vs flexible (restraints only on the hydrogen bonds of the Watson-Crick base pairs, for more details on the restraints used refer to the methods) and \kp or KCl vs Uniform positive background (UB+). On the GpG the RNA atoms were not frozen, instead the RMSD with respect to an equivalent RNA duplex fragment was restrained to 0 nm.
  }
  \label{simsystems}
\begin{tabular}{  c  C{3.5cm} | C{1.5cm}  C{0.7cm} C{0.7cm}  C{0.7cm}  C{2.0cm} C{1.5cm} C{2.0cm} }
\hline
\multicolumn{2}{c |}{\multirow{2}{*}{System Description}} & \multicolumn{1}{C{1.5cm}}{\multirow{2}{\linewidth}{Num. of atoms}} & \multicolumn{3}{c}{Num. of ions} &{\multirow{2}{\linewidth}{Num. of waters}} & {\multirow{2}{\linewidth}{V$_{box}$(nm$^3$)}} & {\multirow{2}{\linewidth}{Sim. time (\textmu s)}} \\
 & & & \multicolumn{1}{c}{\mg} & \kp & \cl & & &  \\
\hline
 \multicolumn{1}{c|}{\multirow{5}{*}{\rot{Nucleosides}}} & Adenosine  & 1545 & 1 & 0 & 0 & 504 & 15.2 & 0.5   \\
 \multicolumn{1}{c|} {} & Cytidine  & 2623 & 1 & 0 & 0 & 864  & 26.6 & 0.5   \\  %
 \multicolumn{1}{c|} {} & Guanosine  & 1480 & 1 & 0 & 0 & 482  & 14.8 & 1.0   \\ 
 \multicolumn{1}{c|} {} & Uridine  & 1698 & 1 & 0 & 0 & 556  & 17.02 & 1.0   \\
 \multicolumn{1}{c|} {} & Rigid GpG & 1814 & 1 & 0 & 0 & 582  & 18.3 & 1.0   \\ \cline{1-9}
 \multicolumn{1}{c|}{\multirow{9}{*}{\rot{$^{GGGG}_{CCCC}$}}} & Rigid $sb$ UB+  & 6577 & 1 & 0 & 0 & 2106  & 66.6 & 9.0   \\
 \multicolumn{1}{c|} {} & Rigid $sb$ \kp & 6569 & 1 & 4 & 0 & 2102  & 66.6 & 9.0   \\
 \multicolumn{1}{c|} {} & Rigid $lb$ \kp  & 73298 & 1 & 4 & 0 & 24345 & 735.1 & 3.0  \\
 \multicolumn{1}{c|} {} & SS Rigid $lb$ \kp & 73256 & 1 & 4 & 0 &  24339  & 735.1 & 3.0   \\ \cline{2-9}
 \multicolumn{1}{c|} {} & Flexible $sb$ UB+ & 6580 & 1 & 0 & 0 & 2107  & 65.9 & 9.0   \\
 \multicolumn{1}{c|} {} & Flexible $sb$ & 6572 & 1 & 4 & 0 & 2103  & 65.8 & 9.0   \\
 \multicolumn{1}{c|} {} & Flexible $sb$ KCl & 6556 & 1 & 8 & 4 & 2095 & 65.8 & 9.0  \\
 \multicolumn{1}{c|} {} & Flexible $lb$ KCl & 33333 & 1 & 24 & 20 & 11010  & 337.4 & 9.0  \\
 \multicolumn{1}{c|} {} & Flexible $lb$ KCl + MgCl & 33323 & 4 & 22 & 24 & 11005 & 336.1 & 9.0  \\ \cline{1-9}
 \multicolumn{1}{c|} {} & $^{GGCC}_{CCGG}$ Flexible $sb$ \kp & 6563 & 1 & 4 & 0 & 2100  & 65.9 & 9.0  \\ [7pt]
\end{tabular}
\end{table*}

The $K_a$ related to individual binding sites were calculated using the equilibrium distributions recovered from the BE-MetaD simulations. 
The simulations were reweighted using the umbrella sampling relationship~\cite{torrie1977umbrella}, combining the last bias from MetaD
 \cite{branduardi2012metad} and the penalty bias potentials added on the different replicas.
 Data from different replicas were combined with the weighted histogram analysis method (WHAM). \cite{kumar1992wham}
This procedure is closely related to the standard way used to analyse BE-MetaD simulations \cite{marinelli2009kinetic} but was here performed in a binless fashion that allows
a large number of restraints to be simultaneously reweighted.
The binding free energy at the standard 1M Mg$^{2+}$ concentration was calculated accounting for the probability of the cation to be found in the bulk region.\cite{gilson1997statistical}
 The bulk region was defined as a spherical shell around the center of mass of the RNA duplex (CoM$_{RNA}$) in which the free-energy profile as a function
 of the Mg$^{2+}$-CoM$_{RNA}$ distance was flat (see SI 3).
The total weight $w_i$  corresponding to the $i$-th binding site was obtained by accumulating the corresponding WHAM weights, and 
the affinities were computed as
$$
  K_a^i=\frac{w_i}{w_{shell}}V_{shell} 
$$
Here $w_{shell}$ is the total weight accumulated in the bulk (shell) region and $V_{shell}$ is its volume.
\mg binding free energies were then defined as $\Delta G_i=-k_B T\log K_a^i$.
Since $K_a$ is expressed in molar units, a positive $\Delta G$ indicates that
at a nominal  \mg concentration of 1 M one would expect the probability of finding a \mg bound to
be smaller than the probability of finding no \mg bound.

The following set of simulations were designed and performed in order to evaluate the effect of the box size and of the ionic composition of the buffer:
(I) a flexible duplex in a large box ($\approx$11000 explicit water molecules) and a buffer of KCl at 0.1 M concentration;
(II) a flexible duplex with the same box and a buffer with KCl and MgCl$_2$ at 0.1 M and 0.02 M concentration respectively;
(III) a flexible duplex with a smaller box ($\approx$2100 explicit water molecules)  and a buffer of KCl at 0.1 M concentration.
In this way, by comparing the affinities and the free energy of the biased \mg against the center of mass of the RNA duplex (see SI 3) we could single out the effect of the having extra \mg in the bulk ((I) vs (II)) and the effects of the box size in the \mg affinity ((I) vs (III)). 

To dissect the contributions to \mg-RNA binding we performed calculations on the following systems using a box with $\approx$2100 explicit water molecules:
(a) a flexible duplex with and without explicit K$^+$;
(b) a rigid duplex with and without explicit K$^+$;
(c) a rigid duplex and two rigid separated strands with sequences GGGG and CCCC in a larger simulation box ($\approx$24000 explicit water molecules). In this latter case the $K^+$ concentration was $\approx$0.01 M.
When not using explicit monovalent ions, $K^+$ was replaced by a uniform positive background (UB+).
This combination of setups allowed for the following factors to be considered: ion competition, RNA flexibility, and RNA hybridization. 
All the simulations followed the same protocols described above.

It is important to consider  that a proper description of the kinetic and thermodynamic behavior of Mg$^{2+}$ cations is very difficult to achieve
without explicitly taking polarization and charge transfer effects into account. \cite{spaangberg2004polarizableff,petrov2011bidentate}
Several models have been introduced to effectively include polarization either using standard force field terms, \cite{allner2012magnesium,merz2013pmemg}
by means of \textit{ad hoc} modified Lennard-Jones potentials, \cite{panteva2015forcefield} or within a Drude model.\cite{Haibo2010drudeions}
We here decided to opt for the parameters developed in ref.\cite{allner2012magnesium} which were optimized to improve Mg$^{2+}$ kinetic behavior in water
and interaction with phosphate.
We already used these parameters in previous applications to model ATP-bound Mg$^{2+}$ and to  describe the effect of  Mg$^{2+}$ on
tertiary contacts in a riboswitch. \cite {Perez2015atp,dipalma2014mgriboswitch}
We notice that a proper balance in \mg-RNA interaction is not granted by available force fields.\cite{panteva2015comparison}
For this reason, we checked the robustness of the reported results by using a reweighting procedure.
We applied a pragmatic correction, adding \emph{a posteriori} a contribution to the interaction between Mg$^{2+}$ and individual binding sites on RNA proportional to a switching
function $V_{correction}=\sum_i \lambda_i (1+(d_i/R_0)^6)^{-1}$.
Here $d_i$ is the distance between \mg and the $i$-th target binding site. 
$R_0$ is a cutoff radius that defines the directly bound state and is chosen to correspond to the barrier separating inner and outer sphere binding.
$\lambda_i$ are Lagrangian multipliers found with an iterative procedure so as to enforce the experimental value of the affinity on individual binding sites.
Affinities calculated on nucleosides as well as $\lambda_i$ and $R_0$ values are reported in SI 4.
The weight used to compute the affinities are then corrected by a factor $e^{-\frac{V_{correction}}{k_BT}}$.
This procedure follows the MaxEnt prescription \cite{pitera2012maxent} 
stating that the minimal correction to a force field so as to enforce the average value of an observable should be proportional to the same observable.
This procedure is expected to provide results comparable to those reported in ref. \cite{panteva2015forcefield}.
The difference between the results with or without these corrections is discussed when appropriate.

\section{Results}

The main output of our simulations is the binding affinity of \mg  on all the possible binding sites in a RNA duplex.
We first report a detailed analysis of the obtained affinities and the correspondence with frequencies from the PDB.
A comparison with the thermodynamic data available for nucleosides and a dinucleoside monophosphate is reported in SI 4.
Then, we show a set of simulations performed in different conditions to dissect the important contributions to \mg-RNA binding.
The reported affinities were calculated by averaging over atoms of the same type in the central bases of the duplex, so as to mitigate terminal effects.
For all the reported quantities we also computed statistical errors. For a quantity whose best estimate is $X$ and the confidence interval is $[X-\Delta_1,X+\Delta_2]$,
the value is reported as $X$$^{+\Delta_2}_{ -\Delta_1}$. Errors are computed by block averaging over 4 blocks without discarding any part of the simulation.
When relevant we also discuss the results obtained by applying a correction that enforces the experimental affinity
on all the binding sites of a nucleoside (MaxEnt correction, see Methods).

\subsection{Mg$^{2+}$ binding on a flexible duplex}

Table \ref{table1} reports the binding affinity for \mg of a flexible RNA duplex in presence of explicit \kp and \cl ions.
Reported results are obtained with the large simulation box ($\approx 11000$ water molecules). 
With our approach one can obtain affinities for both inner and outer sphere binding on all the possible binding sites.
For the sake of clarity we define as outer sphere binding any state in which an explicit water molecule is between \mg and a electronegative donor.

Affinity for inner sphere binding is dominated by the phosphates, with a preference for the strand composed of guanines.
We observed a significant preference for direct binding on G-O2P with respect to the G-O1P.
Nitrogens that are involved in base pairing never formed direct contacts with \mg. Affinity of C-O2 was extremely low, being surpassed by the O2' in the sugar moiety.
The only atoms in the nucleobase displaying significant affinities were G-O6 and G-N7.
All these observations are in striking agreement with interaction frequencies observed in the PDB taken from ref. \cite{bujnicki2015pdbfreq} that are also reported in Table \ref{table1}.
The only exception is the inversion in the binding free energies of C-O2 and O2'. Our underestimation of the affinity
of C-O2 might be biased by our choice to simulate a RNA duplex. Indeed, affinity of C-O2 is expected to be increased when cytosine is not involved in a canonical base pair.
Interestingly, in a simulation performed on a isolated nucleoside (see SI 4) the affinity of C-O2 was significantly larger.

When looking at outer sphere binding, both nucleobase and phosphate backbone contribute to the overall affinity (see Table \ref{table1}).
Also in this case, there is a preference for the strand composed of guanines, and binding on G-O2P is more favorable with respect to binding on G-O1P.
The affinity of G-O6 and G-N7 is comparable to the affinity of phosphates. Also the sugar oxygens have a relatively large affinity. 
These observations agree with the interaction frequencies observed in the PDB.
The only two exceptions are related to O5' and O3', for which the reported frequency is low and 
to the G-O6 and G-N7 affinities which are in the inverse order but within error of each other. 
The former discrepancy could be related to the fact that, at variance with our approach, in the reported experimental frequency the outer sphere binding with phosphates were excluded from the count on the O5' and O3' interaction frequencies. 
The latter discrepancy could be related to the sequence we choose to sample.
It must be also noticed that binding of \mg on G-N7 can happen in a large variety of equivalent and consequently
non-comparable structural contexts.
Moreover, the difference in the reported experimental frequencies is very small.  

One might be tempted to convert the observed frequencies into binding free energies that can be quantitatively compared with our results.
Even though the correlation is good ($R^2 = 0.61$ for inner sphere and $R^2 = 0.67$ for outer sphere binding), indicating that the ranking is consistent, the slope of the fitting line is very far from unity (see SI 5). This might be due to imbalance in the force field \cite{panteva2015comparison}
as well as to the fact that PDB distributions are not necessarily representative of the canonical ensemble.
Additionally, it is not clear how much these frequencies could be used to anticipate location of \mg ions in solution.

Remarkably, our calculation can recapitulate the most important trends observed in experimental frequencies, namely: preference for G with respect to C;
preference for major-groove with respect to minor groove; relative preference between all the relevant binding sites.

\begin{table}[ht]\footnotesize
  \caption{Calculated \mg affinities on a duplex and PDB frequencies from ref. \cite{bujnicki2015pdbfreq}. Frequencies for
  sugar and phosphate moieties were reported independently of the base identity.
  }
  \label{table1}
\begin{tabular}{  c  c  c | R{1.0cm}  R{1.0cm} | R{1.0cm}  R{1.0cm} }
\hline
 \multicolumn{3}{c |}{{\multirow{2}{*}{Binding sites}}} & \multicolumn{2}{ c |}{Inner sphere} & \multicolumn{2}{c }{Outer sphere} \\ \cline{4-7}
 & & & \multicolumn{1}{C{1.0cm}}{$\Delta G^{inner}$ (kJ/mol)} & F$_{PDB}^{inner}$ & \multicolumn{1}{C{1.0cm}}{$\Delta G^{outer}$ (kJ/mol)} & \multicolumn{1}{C{1.0cm}}{F$_{PDB}^{outer}$} \\
 \hline
 \multicolumn{1}{c|}{\multirow{9}{*}{\rot{Bases}}} &  \multirow{6}{*}{G} & N1   &    --- &  ---  & 7.6 &  0.22 \\
\multicolumn{1}{c|} {}   &   &  N2   &    --- &  ---  &  4.0 &  0.11 \\
\multicolumn{1}{c|} {}   &   &  N3   &    --- & 0.002 &  5.6 &  0.12 \\
\multicolumn{1}{c|} {}   &   &  N9   &    --- &  ---  &  0.6 &  0.01 \\
\multicolumn{1}{c|} {}   &   &  N7   &   25.2 & 1.35  & -10.9 &  3.62 \\
\multicolumn{1}{c|} {}   &   &  O6   &   22.4 & 1.45  & -10.3 &  3.84 \\ \cline{2-7}
\multicolumn{1}{c|} {}   & \multirow{3}{*}{C}  &  N1   &    --- &  ---  & 15.2 &  0.008 \\
\multicolumn{1}{c|} {}   &   &  N3   &    --- & 0.01  & 21.8 &  0.33 \\
\multicolumn{1}{c|} {}   &   &  O2   &   59.9 & 0.14  &  4.6 &  0.36 \\ \cline{1-7}
\multicolumn{1}{c|} {\multirow{8}{*}{\rot{Sugar}}}   & \multirow{4}{*}{G}  &  O2'  &    33.4 & 0.07  &  0.4 &  0.54 \\
\multicolumn{1}{c|} {}   &   &  O3'  &    --- & 0.04  & -5.3 &  0.55 \\
\multicolumn{1}{c|} {}   &   &  O4'  &    --- & 0.004 &  5.7 &  0.07 \\
\multicolumn{1}{c|} {}   &   &  O5'  &    --- & 0.04  & -8.6 &  0.61 \\ \cline{2-7}
\multicolumn{1}{c|} {}   & \multirow{4}{*}{C}  &  O2'  &   34.4 & 0.07  &  1.0 &  0.54 \\
\multicolumn{1}{c|} {}   &   &  O3'  &    --- & 0.04  & -3.9 &  0.55 \\
\multicolumn{1}{c|} {}   &   &  O4'  &    --- & 0.004 &  4.7 &  0.07 \\
\multicolumn{1}{c|} {}   &   &  O5'  &    --- & 0.04  & -4.3 &  0.61 \\ \cline{1-7}
\multicolumn{1}{c|} {\multirow{4}{*}{\rot{Phosp.}}}  & \multirow{2}{*}{G}  &  O1P  &  -43.5 & 4.19  & -8.4 &  1.91 \\
\multicolumn{1}{c|} {}   &   &  O2P  &  -48.1 & 4.99  & -10.5 &  2.78 \\ \cline{2-7}
\multicolumn{1}{c|} {}   & \multirow{2}{*}{C}  &  O1P  &  -22.8 & 4.19  & -7.0 &  1.91 \\
\multicolumn{1}{c|} {}   &   &  O2P  &  -28.2 & 4.99  & -6.6 &  2.78 \\
\hline
\end{tabular}
\end{table}

\subsection{Dissecting contributions to affinity}

We then repeated the calculations in several different conditions with the aim of dissecting all the contributions to RNA-\mg binding affinity.
Simulations with explicit \kp ions were compared with equivalent simulation using a uniform positive background (UB+) in order to account for ion competition effects.
In the same spirit, simulations with flexible RNA were compared against equivalent ones with rigid RNA, in order to account for flexibility and conformational entropy effects.
These simulations are performed with a box containing approximately 2100 water molecule, which is large enough to observe a clearly flat free-energy profile
as a function of the distance of the \mg from the CoM$_{RNA}$. Profiles obtained with control simulations are shown
in SI 3.
The hybridization effects on \mg affinity can be clarified by comparing simulations done on single stranded (ssRNA) and double stranded RNA (dsRNA).
The conditions for each set of simulation are discussed in detail in the methods.

\subsubsection{Ion competition}

\begin{figure}[h]
   \includegraphics[width=\columnwidth]{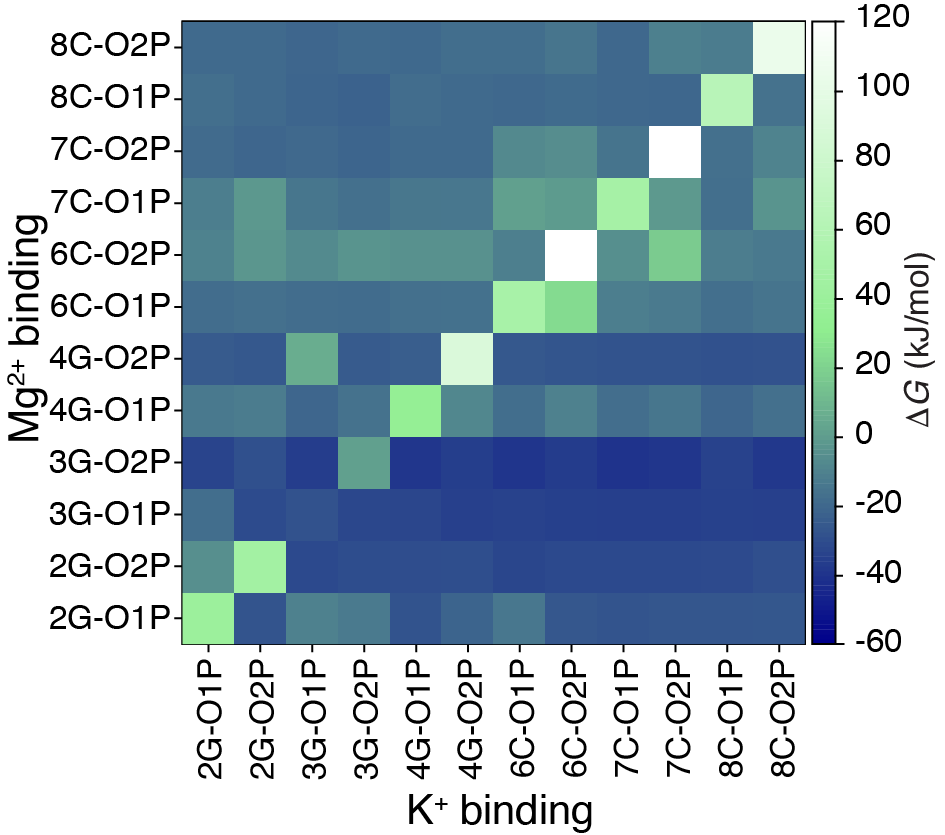}
   \caption{Conditional \mg affinity for phosphate oxygens upon \kp binding.
   Each square represents in a color scale the affinity of \mg on a specific binding site (vertical axis)
   when a \kp is close to another binding site (horizontal axis).
   Since some of the sites are rarely occupied by \kp, statistical errors for those sites
   are large.
   An equivalent matrix including all the \mg binding sites is reported in SI 6.
   }
   \label{mgk_onlyP}
\end{figure}

\begin{figure*}[h]
   \centering
   \includegraphics[width=14.5cm,height=14.5cm]{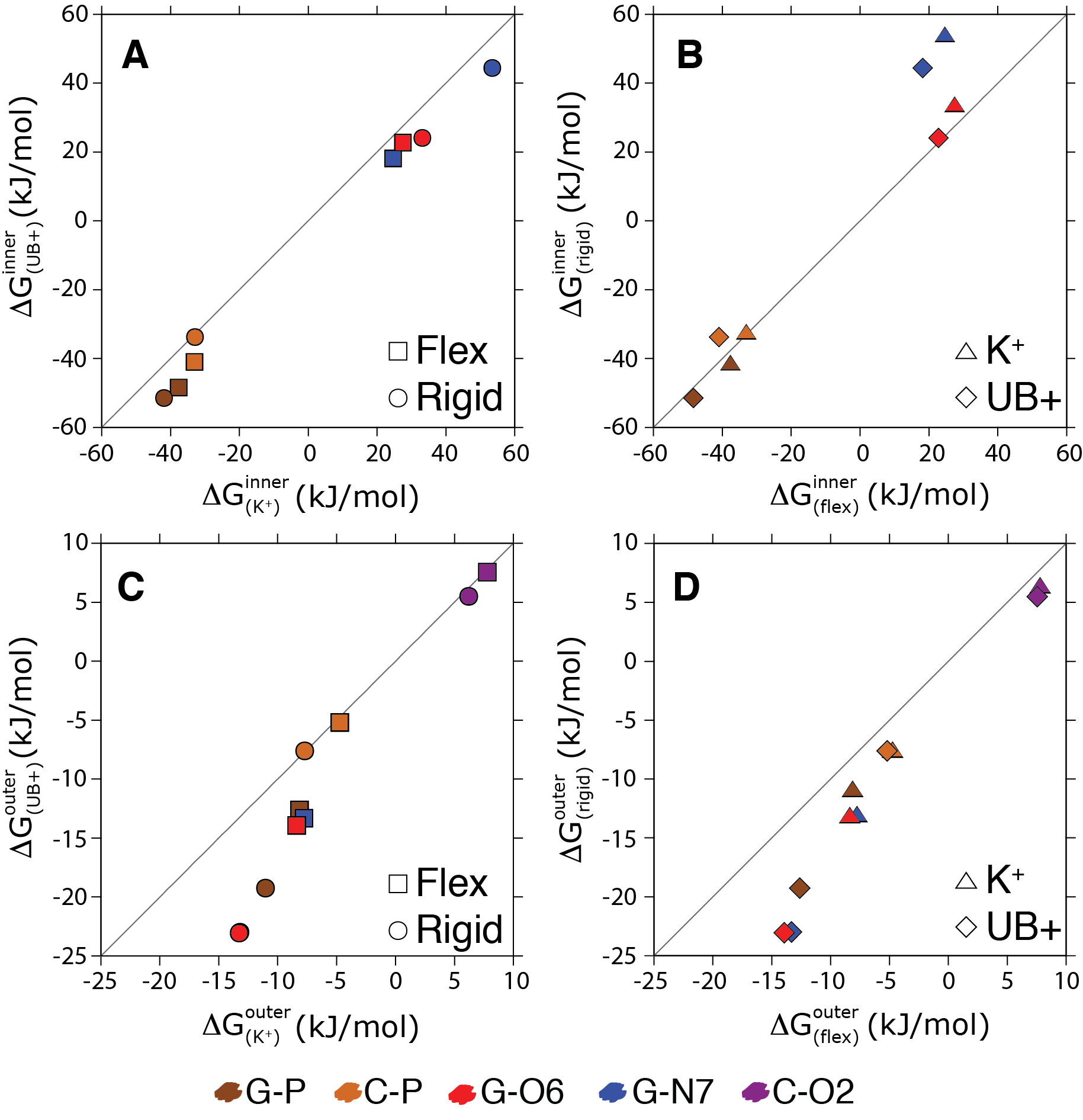}
   \caption{Specific \mg binding affinities on a $^\text{GGGG}_\text{CCCC}$ duplex under different simulation conditions.
   The affinities were obtained in a flexible and rigid duplex both with explicit \kp ions and without, thus with a uniform positive background (UB+).
   Plots A and C show the effect of ion competition (\kp \textit{vs} UB+) for inner and outer sphere \mg binding respectively. 
   Plots B and D show the effect of flexibility (flex \textit{vs} rigid) for inner and outer sphere \mg binding respectively.
   }
   \label{flexioncomp}
\end{figure*}

We first use our simulations to quantify how much the competition with \kp influences the RNA affinity for \mg.
To this aim we compared the affinities on individual binding sites using either an uniform positive background (UB+)
or explicit K$^+$ ions. To avoid complications related to the interaction of counterions and coions,
we only added 4 monovalent cations so as to neutralize the system.
Inner sphere binding free energies reported in Figure \ref{flexioncomp}A shows the effect of explicit \kp ions on \mg-RNA affinity.
Here it is possible to appreciate that
competition of K$^+$ ions decreases the overall Mg$^{2+}$ affinity, both when RNA is kept rigid and when it is modeled as flexible.
The change in the total
binding free energy is quantified as 10.5 $^{+1.0}_{-0.7}$ kJ/mol for flexible RNA and as 9.5 $^{+0.7}_{-0.6}$ kJ/mol for rigid RNA.
The effect of the presence of K$^+$ ions on the affinity  can be also rationalized
by measuring the Mg$^{2+}$ affinity on individual sites when there is a K$^+$ ion in proximity of the same site. 
Results are reported in figure \ref{mgk_onlyP} and are consistent with the fact that the decreased affinity is
an effect of the competition between the two species for the same binding site.
We notice that the effect of competition is local and propagates in a few case to the nearest neighbor binding sites,
suggesting that also a short model duplex can be used to quantify ion competition.

The binding free energies for the indirectly bound ions are also reported (Figure \ref{flexioncomp}C) and follow a similar trend
being reduced by 4.9 $^{+0.03}_{-0.03}$ kJ/mol for flexible RNA and 9.3 $^{+0.06}_{-0.06}$ kJ/mol for rigid RNA.
Errors are much smaller here since the number of binding and unbinding events
is significantly larger in the case of outer-sphere binding.

Equivalent data obtained including MaxEnt corrections are shown in SI 8.
The changes in binding free energies due to flexibility are
are not  affected by the corrections on the \mg parametrization, indicating that these results are robust with respect to
the imbalance between binding on nucleobases and phosphates observed in the original force field.

\subsubsection{RNA flexibility}

We then compared the results obtained with a rigid RNA molecule with those obtained with a flexible one.
The flexible RNA molecule had minor restraints so as to conserve its secondary structure, but still could undergo significant local
deformations.
Flexibility effects on the affinity of the inner sphere bound ions are reported in Figure \ref{flexioncomp}B.
The effect of flexibility is not trivial.
In the system where cations are explicitly included, the affinity of Mg$^{2+}$ on flexible RNA
is decreased by 3.9 $^{+0.9}_{-0.7}$ kJ/mol with respect to rigid RNA. 
In the system where cations are replaced with a UB+ the affinity of Mg$^{2+}$ on flexible RNA with respect to rigid
is decreased by 2.9 $^{+0.8}_{-0.6}$ kJ/mol.

The values of affinity for the indirectly bound ions (Figure \ref{flexioncomp}D ) follow the same direction,
decreasing by 4.3 $^{+0.05}_{-0.05}$ kJ/mol in the simulation with explicit K$^+$ ions
and 8.7 $^{+0.04}_{-0.04}$ kJ/mol to the one with a UB+.

Also these values are barely affected by the corrections on the \mg parametrization, indicating that these results are robust with respect to
the imbalance between the binding affinity of nucleobases and phosphates observed in the original force field. Equivalent data
obtained including MaxEnt corrections are shown in SI 8.

By dissecting the contribution of the individual binding sites to the overall affinity, it can be seen that
the central phosphate of the guanine (G-P), which contributes most to the overall affinity, has
a greater affinity for Mg$^{2+}$ when RNA is kept frozen. This is true for all the three G-Ps.

Interestingly, the affinity on the nucleobase binding sites located in major groove (G-O6 and G-N7) is affected by flexibility with an opposite trend.
The lower affinity in the ideal rigid structure 
indicates that the duplex should undergo slight rearrangements so as to bind \mg on the major groove.
Figures SI 9 and SI 10 show the conditional probability distributions of all RNA backbone dihedrals consequent to \mg direct binding for all phosphates. No significant rearrangement can be appreciated on the backbone dihedrals. We notice however that a very small repositioning of the
phosphates could lead to a significant change in the electrostatic interaction with \mg that would explain the observed differences.
Therefore, the structural integrity of the duplex was maintained even when \mg was directly bound to RNA.
It is also relevant to say that for $\approx$1\% of the simulation time flexible RNA underwent a reversible transitions to ladder-like structures. \cite{banas2010ladder}
 This is consistent with what has been observed in recent simulations of restrained RNA duplexes. \cite{bergonzo2015highly}
 Reversibility was checked by monitoring the continuous trajectories so as to avoid false transitions to be observed just due to replica exchanges.
 To assess the impact of these structures on \mg binding,
we recomputed all the affinities by excluding all the snapshots where at least one of the glycosidic torsions was in the range $(-90^\circ,0^\circ)$,
which corresponds to the \emph{high anti} conformation observed in ladder-like structures. Only the affinities for O5' at the 5' termini were affected.
All the other affinities where within the statistical error from the calculation including all the data,
indicating that transitions to ladder-like structures is not correlated with \mg binding.
Since only the affinities for non-terminal nucleotides are discussed in this work, the presence of ladder-like structure does not affect the reported results.

\begin{figure}[ht]
   \includegraphics{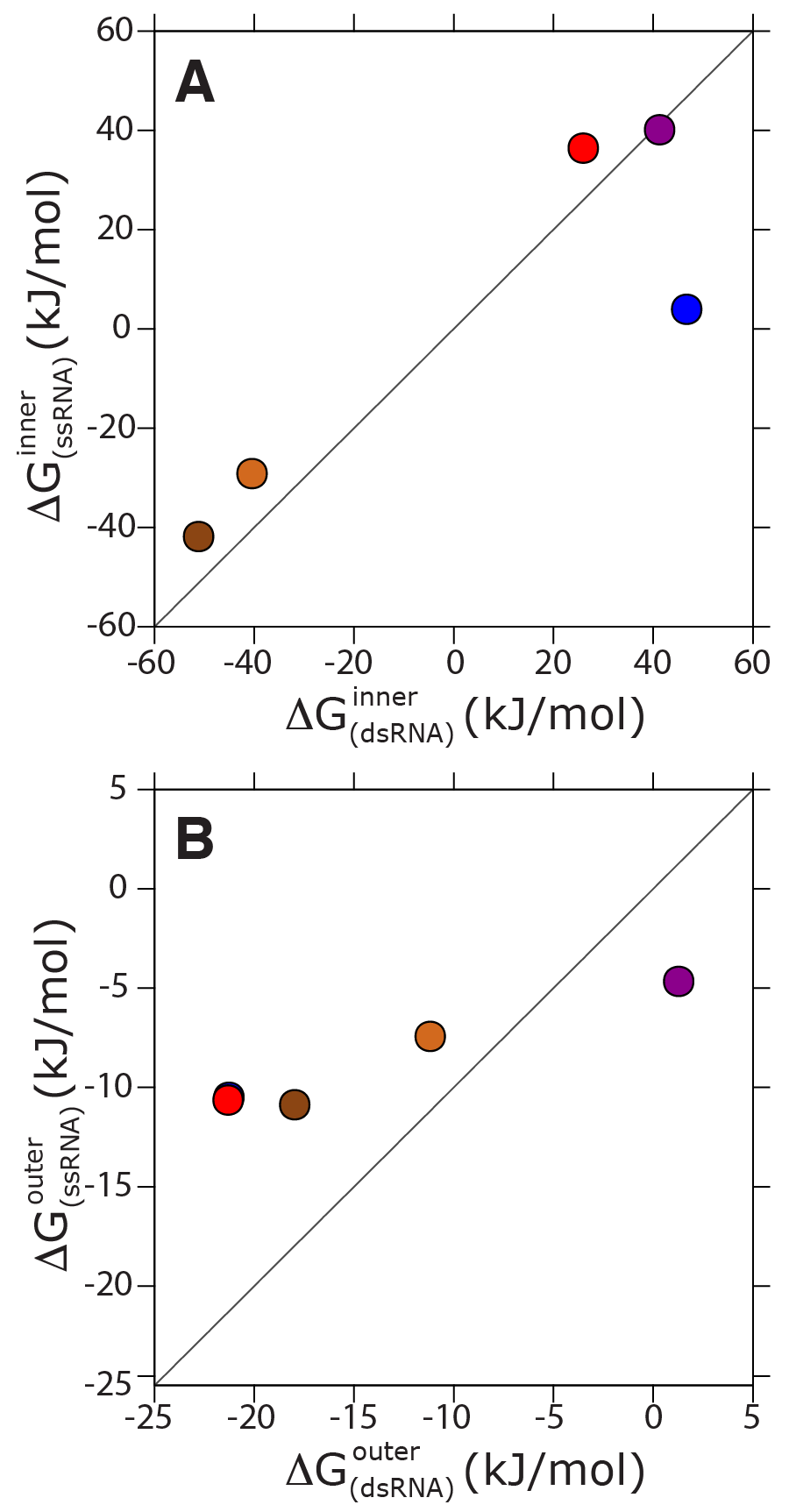}
   \caption{\mg binding affinities on a dsRNA vs ssRNA. Plot (A) shows inner sphere binding and plot (B) outer sphere. 
    The points are color coded according to Figure \ref{flexioncomp}.
   }
   \label{hybridization}
\end{figure}

\subsubsection{Duplex hybridization}

Finally, we compare the affinity of Mg$^{2+}$ with a  double stranded RNA (dsRNA) against the one with a pair of single stranded RNAs
with the same sequence.
The calculations were performed for both systems in a box that was large enough to contain
the two separated strands and in identical ionic conditions. Also these simulations were performed using neutralizing cations only,
so that the affinities reported for the duplex presented in this Section corresponded to a lower ionic strength in comparison
with those presented above.
In all these simulations, RNA was kept rigid.
Indeed, sampling all the conformations available for a ssRNA is a formidable task \cite{gilley2015enhanced,bergonzo2014multidimensional} 
and would have made virtually impossible to obtain converged values for the binding affinities. 
Moreover, the capability of current force fields to 
correctly reproduce the conformational ensembles of ssRNA has been questioned. \cite{bergonzo2015highly,condon2015stacking,Bottaro2016RNAfolding,gilley2016empiricalcorr}
To allow for the affinities to be comparable, it was necessary to treat also the dsRNA as rigid.
Affinities for inner sphere binding are reported in Figure \ref{hybridization}A.
Overall the affinity in the dsRNA was larger
indicating that hybridization and Mg$^{2+}$ binding act cooperatively.
In other words,
when a Mg$^{2+}$ ion is interacting with RNA, the hybridization free energy is expected to be
decreased by 9$^{+1.4}_{-0.9}$ kJ/mol, further stabilizing the duplex. 
We also notice that the overall affinity on the dsRNA is dominated by direct interactions with the phosphate.
However, it is interesting to see this effect on individual binding sites.
In Figure \ref{hybridization} it can be seen that the affinity with the G-O6 is affected by hybridization in the opposite manner,
so that affinity in the ssRNA is enhanced. This is consistent with the fact that electronegative atoms in the base are more
accessible to divalent ions. However, since the contribution of bases to the overall affinity on the dsRNA is negligible with respect 
the contribution of phosphates, this effect is not visible in the overall affinity.

It is also possible to compare the affinity of Mg$^{2+}$ ions which are directly bound with that of ions that 
are indirectly bound.
As it can be seen in Figure \ref{hybridization}B, hybridization increases the stability of indirect binding sites as well, by 9.7 $^{+0.1}_{-0.1}$ kJ/mol.

\section{Discussion}

In this paper we present extensive molecular dynamics simulations investigating the binding affinity of \mg on all the possible binding sites in a RNA duplex.
Calculations are performed using state-of-the-art force fields and a modified version of bias-exchange metadynamics.

The enhanced sampling method we used is an improved version of the  bias-exchange metadynamics (BE-MetaD) procedure designed for this application.
 At variance with the original approach, we here added different penalizing potentials in each replica
to avoid  binding events that would trap Mg$^{2+}$ in undesired positions.
This procedure was necessary here to achieve converged affinities.
The idea of forbidding different events in different replicas can be straightforwardly generalized
to cases where one wants to study competing rare events under the same conditions, and could provide a significant
improvement in the applicability of BE-MetaD to the description of complex free-energy landscapes.
To help reproducibility of our results and application of the procedure to different systems, we included sample input files in SI 2.

Overall, our procedure provides statistically converged affinities for the modeled systems.
To further check convergence, we applied the same protocol on a symmetric duplex (see SI 11).
This check would show if the reported statistical error were underestimated, since a difference in the affinity between the sides of a perfectly symmetric duplex
could only arise from statistical uncertainty.
Our results rely on a few methodological choices and approximations that we here discuss in detail.
It is known that the current RNA force fields may not properly describe 
unstructured single-stranded oligonucleotides. \cite{bergonzo2015highly,condon2015stacking,Bottaro2016RNAfolding}
In this work we focused the investigation on \mbox{dsRNAs} and on rigid ssRNA. 
With respect to the \mg ion itself we assessed our chosen force field by testing the effect of an \emph{a posteriori} adjustment of its interactions with RNA
so as to enforce binding affinities to be in agreement with potentiometric titration experiments on nucleosides. 
Similarly to a recently published parametrization,\cite{panteva2015forcefield} this procedure
did not affect the ion-water and ion-ion interactions.
However, the qualitative consistency between the results with and without the corrections indicates robustness  with respect to \mg force field choice.
Another point to consider is that in our simulations we assumed a single \mg ion binding to RNA at any time, implying an infinite \mg dilution.
To verify the effect of the neglecting extra \mg in the buffer we performed an extra control simulation including an appropriate MgCl$_2$ buffer.
Our results on the competition between \kp and \mg suggest that it would be very difficult for multiple \mg ions to bind
on the same site. However, the double charge of \mg could allow for longer range interactions, affecting the affinity of the neighboring binding sites.
One in principle should thus simulate a replica corresponding to each pair of potentially cooperative binding sites. Ideally, this could be done after
an initial screening where the most important binding sites have been identified.
Additionally, since our simulation box only had enough ions to counterbalance the negative charge of RNA backbone,
our results did not include the effect of anions.
To verify the effect of the neglecting  \cl in the buffer we performed an extra control simulation including an appropriate KCl buffer.
We notice that in all these simulations the number of ions rather than their chemical potential is kept constant.
This limitation could be overcome using a very large simulation box and an extra potential to control ionic strength
in the spirit of ref. \cite{Perego2015chempotential}.
Additionally, the reported control simulations performed with a large box allow the possible artifacts related to box size to
be assessed. Here the \mg binding affinity might be affected by a larger effective 
cation concentration in the vicinity of the RNA (see SI 3). However, our results show that relative binding affinities
are virtually independent on this effect.
Finally the reported results were obtained using a single RNA sequence in a A-form helix.
Sequence and structure dependent effects will be the subject of a further investigation.

Our results show that 
the overall affinity of the inner sphere (direct) bound \mg cations on a RNA duplex is largely dominated by the interaction with phosphates. On the contrary, outer sphere (indirect)
bound \mg cations interacts more strongly with the nucleobases.
We observe that there is a significant preference for inner \mg binding on the guanines with respect to cytosines. Interestingly,
this is consistent in all our simulations including the ones performed with two separated single strands in the A-helix.
This suggests that indirect contacts with guanine N7 might provide extra stabilization.
Additionally, we see an overall preference between the three moieties of
a nucleotide in the following order:  phosphate > bases > sugar.
The affinities on specific binding atoms follows the trend O2P > O1P > G-O6 > G-N7 > sugar hydroxyls > C-O2. 
This trend is not the same when considering the outer sphere contacts, where it is changed to G-N7 $\approx$ G-O6 $\approx$ O2P > O1P > sugar hydroxyls. 
Our procedure captures the experimental trends observed in the PDB binding frequencies both for inner and outer sphere binding.
It must be noticed that the PDB survey reported in ref. \cite{bujnicki2015pdbfreq} 
considers binding with a variety of RNA motifs.
However, the most common RNA motif present in the PDB is the A-helix, which is the same motif addressed by our study.
Although \mg is expected to mostly bind on specific structures and to stabilize tertiary contacts,
the comparison of our study with the discussed PDB analysis indirectly confirms that the patterns of the electrostatic field in the neighborhood of a helix
are representative for the general trend observed in structured RNAs.
It must be also observed that the interpretation of primary X-ray data is not trivial and the assignment of 
many of the reported density peaks to \mg ions have been recently challenged. \cite{Leonarski2016csdsurvey}
However, whereas these errors could affect the interpretation of specific important structures, we expect the overall statistics to be reliable.
Moreover, we notice that, although the ranking are correctly reproduced by our calculations, the reported frequencies
are not proportional to $e^{-\frac{ \Delta G_{Mg^{2+}}^i }{k_BT} }$.
It is not clear whether the frequencies from the PDB can be assumed as representative of a Boltzmann ensemble. 
This discrepancy could also be related to an imbalance inherent to the force field in the
description of interactions of \mg with phosphates and bases which has also been reported in refs. \cite{panteva2015comparison,panteva2015forcefield}.
Ions in MD simulations are usually described by charged Van der Waals spheres. Although this model has proven to be very useful, 
its accuracy is still debated. The main source of doubt comes from the fact that usual MD does not explicitly includes polarizations effects.
 
Our procedure also allowed for the dissection of the effect of ion competition, RNA flexibility and RNA hybridization to \mg affinity.
We found that, for both inner and outer sphere binding,
ion competition and RNA flexibility reduce \mg overall binding affinity while hybridization increases it.

The effect of ion competition on the inner sphere binding was independent of solute flexibility
and amounted to $\approx$ 10 kJ/mol.
The same was true for outer sphere binding on a rigid RNA. However, the effect for outer sphere binding on flexible RNA
was significantly smaller ($\approx$ 4 kJ/mol). This indicates that local rearrangements that are possible in the flexible RNA
could compensate for the repulsion between the cations.

Interestingly, RNA flexibility decreased its affinity for  \mg. We recall that the total affinity is dominated by the phosphates. Affinity on the nucleobases
was on the other hand increased by flexibility.
We argue that the flexibility of RNA might affect binding affinity in two opposite ways.
First, the enthalpic contribution to the affinity could be increased by RNA flexibility when local rearrangements
lead to more favorable RNA-\mg interactions.
On the other hand, binding of RNA with \mg  constrains RNA leading to a loss in its conformational entropy. \cite{bergonzo2016divalent}
Interestingly, it has been recently suggested that multivalent cations make RNA helices more rigid. \cite{Drozdetski2016entropy}
The simulation of rigid RNA allowed us to explicitly ignore changes in the RNA conformational entropy.
We observe that nucleobases are significantly constrained by Watson-Crick pairing and require a local rearrangement so as to bind \mg.
Conversely, phosphates are accessible for \mg binding even in a rigid RNA model. 
We  hypothesize that in the case of \mg binding on phosphates the second effect dominates over the first effect
leading to a decreased affinity in the flexible model.

During the revision process of this paper, an alternative approach was proposed to find \mg binding sites
with large affinity using a grand-canonical Monte Carlo scheme. \cite{MacKerell2016characterization}
Since the approach presented here addresses the same problem in an orthogonal direction, the two schemes might be combined
so as to allow for an even more efficient simulation protocol.

We here presented a computational approach to the detailed characterization of \mg-RNA binding.
To this aim, we introduced a modified version of bias-exchange metadynamics.
Our results reproduce statistics observed in structural databases and allow for a dissection of the
most important contribution to \mg-RNA interactions, shading a new light on the interplay between
RNA flexibility and binding with divalent cations.
We foresee the application of our computational approach to  the characterization of
\mg binding sites on a repertoire of RNA motifs and sequences.
Although there is still controversy regarding \mg ion parameters and the accuracy of the simple models employed the trends in the binding
affinities would likely be consistent. Additionally, the introduced procedure could be used as a benchmark to compare several models.
More generally, our procedure could be used to trustfully quantify 
affinities of ions or small ligands when multiple competing binding sites have to be simultaneously assessed.

\begin{acknowledgement}
Pascal Auffinger, Sandro Bottaro, Janusz Bujnicki, Kathleen Hall and Stefano Piana-Agostinelli are acknowledged for carefully reading the manuscript and
providing useful suggestions. Alejandro Gil-Ley and Simon Poblete-Fuentes are acknowledged for useful discussions.
The research leading to these results has received funding from the European Research Council under the European Union's Seventh Framework Programme (FP/2007-2013) / ERC Grant Agreement n. 306662, S-RNA-S.

\end{acknowledgement}

\providecommand{\latin}[1]{#1}
\providecommand*\mcitethebibliography{\thebibliography}
\csname @ifundefined\endcsname{endmcitethebibliography}
  {\let\endmcitethebibliography\endthebibliography}{}

\end{document}